\documentclass[twocolumn,aps,prc,showpacs,superscriptaddress]{revtex4}
\usepackage{epsfig}
\usepackage{color}
\begin{document}
\title{Event mixing does not reproduce single particle acceptance convolutions for nonuniform pseudorapidity distributions}
\author{Lingshan Xu}
\affiliation{Department of Physics, Purdue University, West Lafayette, Indiana 47907, USA}
\author{Chin-Hao Chen}
\affiliation{RIKEN BNL Research Center, Brookhaven National Laboratory, Upton, New York 11973, USA}
\author{Fuqiang Wang}
\affiliation{Department of Physics, Purdue University, West Lafayette, Indiana 47907, USA}
\begin{abstract}
We point out that the mixed-event method for two-particle acceptance correction, widely used in particle correlation measurements at RHIC and LHC, is wrong in cases where the single particle pseudorapidity distribution is significantly nonuniform. The correct acceptance should be the convolution of two single-particle efficiency$\times$acceptance functions. The error of the mixed-event method, which guarantees a uniform $\Delta\eta$ two-particle combinatorial density, is, however, small in correlation analyses where the two particles are integrated over an extended pseudorapidity $\eta$ range. With one particle fixed in $\eta$ and the right acceptance correction, the background-subtracted correlated pair density may reveal not only a short-range but also a long-range $\Delta\eta$ dependence. This has important physics implication, and may provide crucial information to disentangle physics mechanisms for the recently observed long-range ridge correlation in asymmetric proton-lead collisions at the LHC.
\end{abstract}
\maketitle
\section{Two-particle acceptance}
Two-particle correlations are a valuable tool to study heavy-ion collisions~\cite{PHOBOSwp,STARwp,PHENIXwp}. Correlation functions are often formed by particle pair density in real events divided by that from mixed-events, where the two particles are taken from different events. With proper normalization the deviation of the correlation function from unity reveals correlations between two particles. This technique applies to situations where all particle pairs are correlated, for example, in Hanbury-Brown and Twiss (HBT) interferometry~\cite{HBT} and anisotropic flow correlations~\cite{Ollitrault}.

One unique case of correlations is jet angular correlations. The study of jet correlations has provided wealth of information about relativistic heavy-ion collisions~\cite{PHOBOSwp,STARwp,PHENIXwp}. The object of interest is a cluster of particles--not all particles in the event--that are correlated due to their common origin of parton fragmentation. The interest is often to find number of correlated particles and their angular distributions. The correlations are often formed in terms of the particle pair azimuthal angle difference, $\Delta\phi$ and pseudorapidity difference, $\Delta\eta$. The mixed-event technique is widely used, not as the uncorrelated baseline as in HBT or anisotropic flow analysis, but to correct for two-particle acceptance. For example, in mid-rapidity region, the mixed-event two-particle density is approximately triangular in $\Delta\eta$, and is used for $\Delta\eta$ acceptance correction after normalized to 100\% at $\Delta\eta=0$. This mixed-event method for two-particle acceptance correction is, however, wrong. The correct two-particle acceptance should be the convolution of two single-particle acceptances (or efficiency functions). It has nothing to do with the {\em measured} single-particle densities, but only depends on detector efficiencies. The mixed-event two-particle density, on the other hand, depends on both the detector efficiencies and the true single-particle densities.

In the widely used mixed-event acceptance correction, the single-particle distributions are mistaken as detection efficiencies, implying that the true distributions (e.g.~the pseudorapidity density $dN/d\eta$) are always uniform. 
By using mixed-event technique, the ``corrected'' two-particle correlation signal in $\Delta\eta$ is guaranteed to be uniform except regions of correlation signals. The truth may, of course, not be uniform as the single-particle $dN/d\eta$ at mid-rapidity is not strictly uniform. 
This, however, does not make a big error for mid-rapidity particles in symmetric collisions, where particle density is nearly uniform in $\eta$~\cite{PHOBOSdndy}. 

The mixed-event acceptance correction could be further from truth for asymmetric collisions, such as proton-lead (p-Pb) collisions, where the single-particle $dN/d\eta$ is nonuniform even at mid-rapidity~\cite{ALICEdndy}. The corrected two-particle correlations will, again, be uniform by construction except the $\Delta\eta$ regions of any correlation signal. 
But this clearly is wrong. It is easy to see in the following simple example. Suppose in a jet-correlation study in $\Delta\eta$, the high-$p_T$ trigger particle is fixed at $\eta=0$ and all other particles are paired with the trigger to form angular correlation function. There will be a peak at $\Delta\eta=0$ due to jet correlations and the underlying background will have the same shape as the underlying event single-particle $dN/d\eta$ distribution. The two-particle acceptance correction to be applied should be the single-particle efficiency as function of $\eta$ (because the trigger particle is always at $\eta=0$). After correction, the signal should be the real jet signal on top of the real background which, in this case, is the true single-particle density which may not be uniform in $\eta$. If the mixed-event $\Delta\eta$ distribution (which is the single-particle $\eta$ distribution in this simple case) is used as the acceptance correction, then the corrected pair density will be completely flat in $\Delta\eta$ except for regions of correlation signal. Clearly, this does not reflect the true physics condition and will be wrong.

In real data analysis, often all pairs of trigger and associated particles are used. In asymmetric collisions, neither the trigger nor the associated particle $\eta$ density may be uniform. Averaging over all trigger and associated particles, the nonuniformities in the single-particle $dN/d\eta$ distributions, however, become a second-order effect in the two-particle density distribution of $dN/d\Delta\eta$. 
To see this, we take the simple example of a measured single-particle density to be linear in $\eta$, 
\begin{equation}
\frac{dN}{d\eta}\propto 1+k\frac{\eta}{\eta_m}\,,
\end{equation}
where $\pm\eta_m$ are the acceptance limits. The combinatorial two-particle density distribution will be
\begin{widetext}
\begin{eqnarray}
\frac{dN}{d\Delta\eta}&\propto&\int_{\eta_1}\int_{\eta_2}\left(1+k\frac{\eta_1}{\eta_m}\right)\left(1+k\frac{\eta_2}{\eta_m}\right)\delta(\eta_2-\eta_1-\Delta\eta)d\eta_1d\eta_2\nonumber\\
&=&\int_{{\rm max}(-\eta_m,-\eta_m-\Delta\eta)}^{{\rm min}(\eta_m,\eta_m-\Delta\eta)}\left(1+k\frac{\eta_1}{\eta_m}\right)\left(1+k\frac{\eta_1+\Delta\eta}{\eta_m}\right)d\eta_1\nonumber\\
&=&\left(2\eta_m-|\Delta\eta|\right)\left[1+\frac{1}{6}k^2\left(2-2\frac{|\Delta\eta|}{\eta_m}-\left(\frac{\Delta\eta}{\eta_m}\right)^2\right)\right]\,.\label{eq}
\end{eqnarray}
\end{widetext}
The effect of nonuniformity is quadratic in $k$. When $k<<1$ which is typically the case for mid-rapidity region, the two-particle density is approximately triangular in $\Delta\eta$. Note that the two-particle density integrated over trigger and associated particle $\eta$ in Eq.~(\ref{eq}) is symmetric with respect to $\Delta\eta=0$ even though those for fixed trigger $\eta$ are all asymmetric. This is because, taking $k>0$ as example, the larger associated particle density at positive $\Delta\eta$ is weighted by the relatively fewer trigger particles at negative $\eta$, and vise versa.

\begin{figure}[hbt]
\begin{center}
\includegraphics[width=0.45\textwidth]{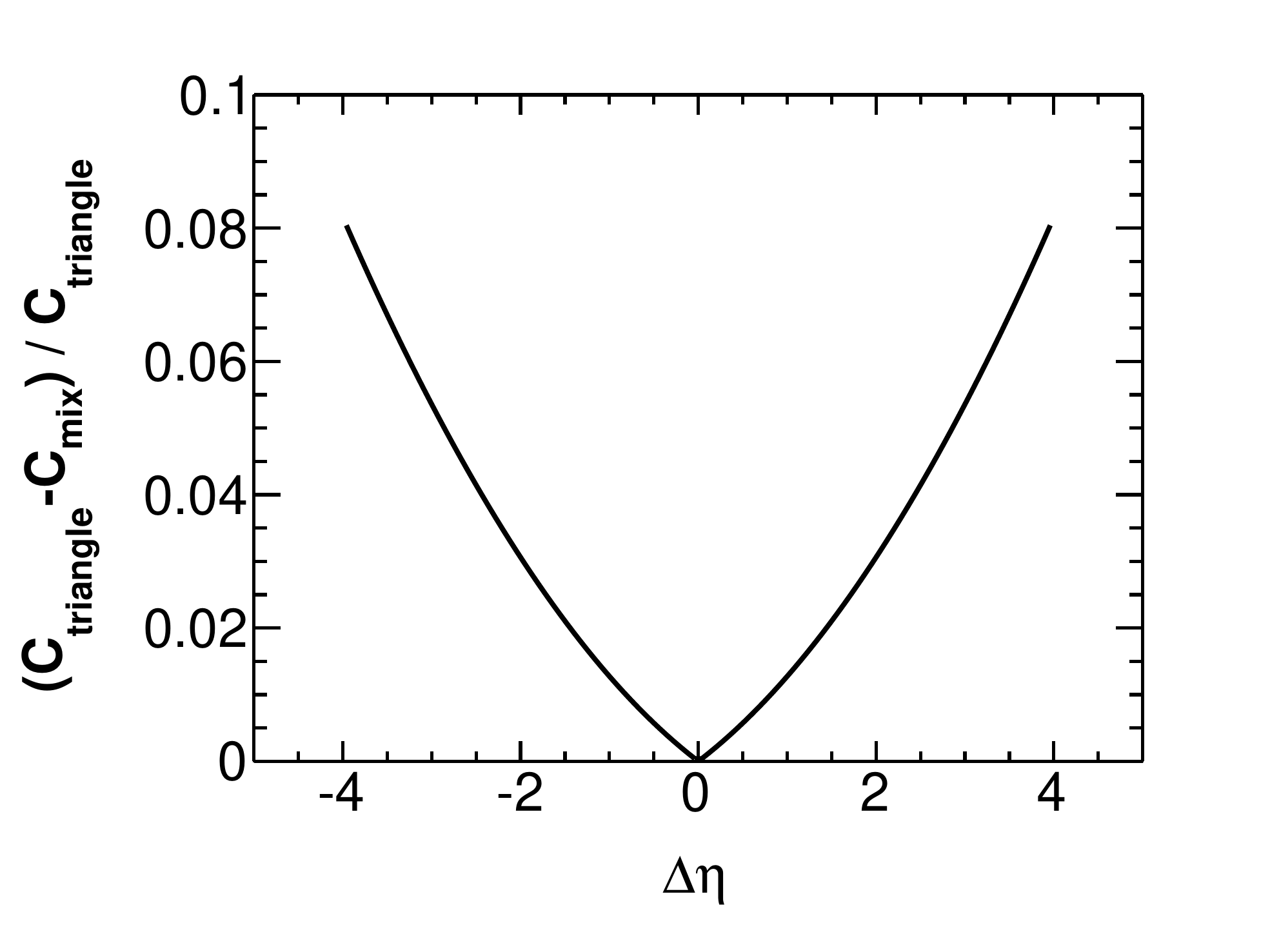}
\end{center}
\caption{Relative error given by Eq.~(\ref{eq2}) that the mixed-event acceptance ($C_{mix}$) makes relative to the true two-particle acceptance ($C_{triangle}$) for a perfect detector (100\% efficiency$\times$acceptance uniform in $\eta$).}
\label{fig}
\end{figure}

In most $\Delta\eta$-$\Delta\phi$ correlation analyses, the mixed-event technique is used for two-particle acceptance correction and, as we have discussed, the final correlated yield is uniform by construction. If the correct acceptance is used from convolution of single-particle efficiencies, the resulting two-particle density does not deviate much from a uniform distribution, as we have shown above. For example, for a perfect single-particle efficiency, the real two-particle acceptance is a triangle. The relative error the mixed-event correction makes by using (the properly normalized) Eq.~(\ref{eq}) is
\begin{equation}
\frac{C_{triangle}-C_{mix}}{C_{triangle}}=\frac{k^2}{6+2k^2}\frac{|\Delta\eta|}{\eta_m}\left(2+\frac{|\Delta\eta|}{\eta_m}\right)\,.\label{eq2}
\end{equation}
For relatively small $k$, the error due to mixed-event two-particle acceptance correction is small. For large $k\approx1/4$ as measured in central 20\% d+Au collisions at RHIC by PHOBOS~\cite{PHOBOSdndy} over a hypothetical acceptance range of $-2<\eta<2$ (i.e.~$\eta_m=2$), Eq.~(\ref{eq2}) gives $\sim3$\% error at $\Delta\eta=\eta_m$ and $\sim8$\% error at $\Delta\eta=2\eta_m$. This is illustrated in Fig.~\ref{fig}.


\section{Physics implication}
We have so far focused on a technical point, that two-particle acceptance correction in $\Delta\eta$ should not be obtained from mixed-events, but rather from convolution of two single-particle efficiency$\times$acceptance functions. Now we want to turn to an important physics implication of this point.

Surprisingly, a strong large-$\Delta\eta$ ($\Delta\eta$ $>$ 2) near-side ($\Delta\phi\approx0$) correlation is observed in p-Pb collisions above a uniform ZYAM (zero yield at minimum~\cite{ZYAM}) background at the LHC~\cite{CMS,ALICE,ATLAS}. This is called the ``ridge'' following the observation in heavy-ion collisions where an elliptic flow modulated background is subtracted~\cite{PRL95,STAR,PHOBOS}. The strength of the ridge in p-Pb is as strong as that in heavy-ion collisions, where the ridge is considered to be primarily a consequence of anisotropic triangular flow. As we have discussed, it is by construction of the mixed-event technique that the observed ridge in p-Pb collisions is uniform in $\Delta\eta$, and it will likely remain uniform even using the correct two-particle acceptance correction because the trigger and associated particles in a wide $\eta$ range are averaged. However, the extra feature of the nonuniform single-particle $\eta$ distribution in p-Pb collisions may be essential to unravel the underlying physics mechanisms for the ridge correlations. In order for any possible $\Delta\eta$ or $\eta$ dependence of the ridge to be observable, one needs to fix the trigger particle $\eta_{trig}$ within a narrow bin and study correlations of associated particles, $dN/d\Delta\eta\approx dN/d\eta_{assoc}$.

There are two leading theoretical models for the physics mechanism of the ridge in small systems. One is hydrodynamics where the initial geometry anisotropy is converted into final-state anisotropic (dominantly elliptic) particle distributions by hydrodynamical evolution~\cite{Piotr}. The anisotropic particle distributions result in an enhanced two-particle density at $\Delta\phi=0$, manifested as the ridge. In this picture, the ridge strength is proportional to the underlying background pair density:
\begin{equation}
\frac{d^2N}{d\Delta\eta d\Delta\phi}=\frac{dN(\Delta\eta)}{d\Delta\eta}\frac{1}{2\pi}\left(1+\sum_{n=1}2V_n\cos n\Delta\phi\right)\,.
\end{equation}
Because the underlying event baseline $dN(\Delta\eta)/d\Delta\eta$ depends on $\Delta\eta$, a measurement of the signal $d^2N/d\Delta\eta d\Delta\phi$ will be informative about the nature of the ridge. Of course, the anisotropic harmonic $V_n$ (product of the two single-particle azimuthal anisotropies, $v_n$) may also depend on $\Delta\eta$. Thus, a measurement of $d^2N/d\Delta\eta d\Delta\eta$ may not uniquely confirm or refute the hydrodynamic explanation, but should provide an important extra information. In this respect, it would be crucial to have predictions of the $\eta$ dependences of $v_2$ and $v_3$ from hydrodynamical calculations. 

The second model is the color glass condensate (CGC)~\cite{Ragu}.
The CGC framework gives particular predictions of the two-gluon production process as a function of $\Delta\eta$. A calculation for Au+Au collisions is given in Ref.~\cite{Dusling} which attributes the ridge to a net effect of the CGC enhanced two-gluon production and the strong collective radial flow. Calculations of the $\Delta\eta$ dependence of the ridge in p-Pb collisions are not yet available but should be extremely valuable. 

Presumably, different physics mechanisms would yield different $\Delta\eta$ dependences of the ridge correlations, as well as different energy dependences. Measurements of the ridge as a function of $\Delta\eta$ at both the LHC and RHIC should, therefore, put stringent constraints on models.

Experimentally, the two-particle acceptance should be taken from the convolution of two single-particle efficiencies. For trigger particles in a fixed narrow $\eta$ range, the two-particle acceptance is approximately the single-particle efficiency$\times$acceptance for associated particles. One obtains the near-side raw yield vs.~$\Delta\eta$ and subtracts the combinatorial background by, for example, the ZYAM procedure~\cite{ZYAM}. The ZYAM magnitude as a function of $\Delta\eta$ may be treated as the underlying background particle pair density. One then examines whether the background-subtracted signal is uniform in $\Delta\eta$, proportional to the background pair $\Delta\eta$ density in the ZYAM region, or of any other shape. 

To examine the ridge yield relative to the combinatorial background, one may not even need to do two-particle acceptance correction. One can simply form the ratio of the pair density from real events to that from mixed-events (i.e.~the original correlation analysis), but with fixed narrow $\eta$ bins for trigger particles, and examine the shape of the correlation function in the ridge region of large $\Delta\eta$. One may not even need mixed-events, but treat the pair density in the ZYAM $\Delta\phi$ region as the combinatorial background and examine the ratio of the pair density in the ridge region at small $\Delta\phi$ to that in the ZYAM region, both of a function of $\Delta\eta$. 


The idea of $\Delta\eta$ dependence of the ridge can be applied to asymmetric heavy-ion collisions, for example, the available Cu+Au collision data from RHIC.

In summary, we point out the misconception in two-particle acceptance correction by the mixed-event technique. The mixed-event acceptance correction is, in principle, wrong and guarantees a uniform $\Delta\eta$ combinatorial pair density (correlation). 
The correct two-particle acceptance should be the convolution of two single-particle efficiency$\times$acceptance functions. The effect of the improper mixed-event correction is, however, small in present experimental data where both the trigger and associated particles are integrated over extended $\eta$ ranges. With a fixed narrow $\eta$ range for trigger particles and the proper acceptance correction, the correlation signal may reveal a $\Delta\eta$ dependence, especially for asymmetric collision systems such as p-Pb collisions recently available at the LHC. A $\Delta\eta$ dependence, if present, could be crucial to discriminate physics models for the observed ridge. 

\section{Acknowledgment}
LX and FW are supported by U.S.~Department of Energy under Grant No.~DE-FG02-88ER40412. CHC is supported by the RIKEN-BNL Memorandum of Understanding on spin physics projects.

\end{document}